\documentstyle[prl,aps,epsf]{revtex}

\newcommand{\beq}{\begin{equation}}
\newcommand{\eeq}{\end{equation}}
\newcommand{\bqn}{\begin{eqnarray}}
\newcommand{\eqn}{\end{eqnarray}}
\newcommand{\bqns}{\begin{eqnarray*}}
\newcommand{\eqns}{\end{eqnarray*}}
\newcommand{\bary}{\begin{array}}
\newcommand{\eary}{\end{array}}

\begin{document}

\twocolumn[\hsize\textwidth\columnwidth\hsize\csname@twocolumnfalse\endcsname

\title{Statistics of Lyapunov exponent in
one-dimensional layered systems}
\author{Pi-Gang Luan and Zhen Ye}
\address{Wave Phenomena Laboratory, Department of Physics,
National Central University, Chung-li, Taiwan 32054}
\date{\today}
\draft\maketitle

\begin{abstract}
Localization of acoustic waves in a one dimensional
water duct containing many randomly distributed air filled
blocks is studied. Both the Lyapunov exponent and its variance
are computed. Their statistical properties are also explored
extensively.
The results reveal that in this system the single parameter scaling
is generally inadequate no matter whether the frequency we consider
is located in a pass band or in a band gap. This contradicts
the earlier observations in an optical case. We compare the
results with two optical cases and give a possible explanation
of the origin of the different behaviors.
 \vspace{8mm}
\end{abstract}]
\section{Introduction}
The fact that the electronic localization in disordered
systems\cite{Anderson} is of wave nature has led to suggestion
that classical waves could be similarly localized in random
systems. The effort in searching for localization of classical
waves such as acoustic and electro-magnetic waves is tremendous.
It has drawn intensive attentions from both theorists
\cite{Baluni,SEGC,Sor,Maradudin93,Ye1,Ye2,SIAM}
and experimentalists\cite{Hodges83,Sch}.
Snice the pioneering work of Anderson et.al.\cite{TAA},
the concepts of universality and scaling have become
important\cite{Sak,Mello,Shapiro,Deych1,Deych2}.
These ideas stem from the insensitivity of the macroscopic
laws to microscopic details; that is, systems or models which
differ from each other on a microscopic level can show identical
macroscopic behavior.
According to the hypothesis of single-parameter
scaling (SPS)\cite{TAA,Deych1,Deych2}, if the localization
behaviors of a 1D disordered system obey SPS, the Lyapunov exponent
or inverse localization that characterizes the degree of
localization will be proportional to its variance.

In a recent study of acoustic (AC) waves propagation in a
one dimensional
randomly layered system\cite{LY} we have found that the statistics
relation between Lyapunov exponent (LE) $\gamma$ and its variance
$\mbox{var}(\gamma)$ do not follow the preditions of
single parameter scaling (SPS).
However, in an earlier study on 1D localization behaviors of
electromagnetic (EM) waves\cite{Deych1}, the author claimed
that the non-universal bahaviors of
LE will disappear and SPS will be restorted
while the randomness of the system exceeds a
critical value.
Since for 1D propagation the AC and EM waves are in fact
mathematically equivalent, i.e., there exists one to one
correspondence between these two kinds of waves, it looks quite
impossible that they can have different localization behaviors.
In order to understand where do the main
differences between these two kinds of models come
from, in this paper we study EM and AC systems simultaneously.
Two EM models and one AC model are studied in this paper.
We find that though the statements made by Deych et.al.
\cite{Deych1} are
correct in their chosen case, however, the
applicability of SPS is more or less based
on the fact that the impedance contrast between the
constituents of the wave media is closed to 1. Without
this restriction then even in the EM systems SPS will not be restored
in the high randomness limit.

This paper is organized as follows. In the next section we
explain the correspondence between EM and AC waves and define the
three models employed in this paper. In Sec. III we we first review
the previous results and then
discuss the numerical results of the three chosen models.
A possible explanation of the origin of the novel properties
of the AC localization are also proposed.
Concluding remarks are given in section IV.
\section{Theory and models}
To begin with, we explain the one to one correspondence of the
1D propagation between AC and EM waves. For simplicity while
without destroying the generality we restrict our discussion to
the monochromatic waves with time dependence $e^{-i\omega t}$.
We also assume that waves are nomally incident on the left
boundary of the media and propagate along the x-axis.
The 1D propagation of AC waves under these assumptions is govened by
\beq
\frac{d}{dx}\left(\frac{1}{\rho}\frac{dp}{dx}\right)
=-\frac{\omega^2 p}{\rho c^2}\label{acwave}
\eeq
where $\rho=\rho(x)$, $p=p(x)$ and $c=c(x)$ represent the mass density,
the pressure, and the phase velocity of the wave in the media,
respectively.
In the special case of layered media considered in \cite{LY},
$\rho$ and $c$ are all constants in a single layer.
Across an interface that separate two layers either $\rho$ or $c$
jump to a diferent value but pressure $p$ and
media vibration velocity
\beq
u=\frac{1}{i\omega\rho}\frac{dp}{dx}
\eeq
must be continuously connected.
The continuity conditions for every interfaces
together with the wave equation (\ref{acwave}) itself determine
the dynamics of the whole system.

Similarly, the equation governing 1D propagation of monochromatic
EM waves can be deduced from Maxwell equation and is written as
\beq
\frac{d}{dx}\left(\frac{1}{\mu}\frac{dE}{dx}\right)
=-\frac{\omega^2\epsilon E}{c_0^2}
=-\frac{\omega^2 E}{\mu c^2}
\label{emwave1}
\eeq
or
\beq
\frac{d}{dx}\left(\frac{1}{\epsilon}\frac{dH}{dx}\right)
=-\frac{\omega^2\mu H}{c_0^2}
=-\frac{\omega^2 H}{\epsilon c^2},
\label{emwave2}
\eeq
where $E=E_{y}$ and $H=H_z$ are the electric and magnetic fields
, $\epsilon$ and $\mu$ stand for the permittivity and
permeability, $c_0$ is the speed of light in vacuum and
$c=c_0/\sqrt{\epsilon\mu}$ is the speed of light in the media.
Comparing Eq.~(\ref{emwave1})
and (\ref{emwave2}) with Eq.~(\ref{acwave}) one can
easily recognize the equivalence between AC and EM models
via the substitutions
\beq
E\rightarrow p,\;\;\;\;H\rightarrow -u,\;\;\;\;
\epsilon\rightarrow 1/\rho c^2,\;\;\;\; \mu
\rightarrow \rho
\eeq
in Eq.~(\ref{emwave1}) or
\beq
H\rightarrow p,\;\;\;\;E\rightarrow u,\;\;\;\;
\epsilon\rightarrow \rho,\;\;\;\; \mu
\rightarrow 1/\rho c^2
\eeq
in Eq.~(\ref{emwave2}).

In spite of these similarities, the two models discussed in
\cite{LY} and \cite{Deych1} indeed have some
different features.
First, in the optical case that described in
Eq.~(\ref{emwave1}), one usually
assumes $\mu=1$. However, in our AC model the corresponding quantity
is $\rho$. The mass density ratio btween water and air is about 775,
a very large value. Second, the phase velocity ratio in our model
is 4.455, whereas in \cite{Deych1} the ratio is 1.0945. Third,
the thickness ratio in our model is 9999/1, much larger than
1/1 that considered in \cite{Deych1}. Fourth, we
randomize the thickness of the water layers (medium with high
phase velocity) and keep the thickness of air layers (medium with
low phase velocity) constant. In \cite{Deych1} they randomize the
thickness of the layers with low phase velocity.

To clarify where do the major differeces come from, we define
three models and study them numerically. The system for each model is a
composite made of two kinds of material $A$ and $B$
with corresponding thickness $a_{j}$ and $b_{j}$ in the
$j$th $A/B$ layer. For simplicity hereafter we assume
$a_j=a$, i.e., all $A$ type layers have the same size.
Any quantity $Q$ in $A$ and $B$ type layers are
denoted as $Q_a$ and $Q_b$ respectively.
\begin{itemize}
  \item Model 1 is an optical model with
  $\epsilon_b/\epsilon_a=2$ and $\mu_b/\mu_a=1$. The thickness ratio is
  $\langle b\rangle/a=1$, where
  $b\in \langle b\rangle[1-\Delta,1+\Delta]$ and $\Delta\in[0,1]$.
  \item Model 2 is also an optical model with
  $\epsilon_b/\epsilon_a=20$ and $\mu_b/\mu_a=1$. The thickness ratio
  is  $\langle b\rangle/a=1$, where
  $b\in \langle b\rangle[1-\Delta,1+\Delta]$ and $\Delta\in[0,1]$.
  \item Model 3 is our previously considered acoustic model
  with $c_b/c_a=4.455$ and $\rho_b/\rho_a=755.2$. The thickness
  ratio is given by $\langle b\rangle/a=9999$, where
  $b\in \langle b\rangle[1-\Delta,1+\Delta]$ and
  $\Delta\in[0,1]$.
\end{itemize}

In a single layer the impedance of AC waves is given by $\rho c$
and the impedance of EM waves is given by $\epsilon c$ or $\mu c$.
Thus we see that in Model 3 the impedance contrast is
about $3365$ and impedance contrast in Model 1 and 2 are about $1.414$
and $4.472$ respectively.

\section{Numerical results}
Before the discussion of the numerical results we first summarize
the relevant results of our previous study\cite{LY}.
There both LE and its variance as functions of frequency were
studied. At low disorders, the variance of LE
inside the gaps is small. Contrast to the optical
case\cite{Deych1}, there are no double maxima inside the gap. With
increasing disorder, double peaks appears inside the allow bands.
When exceeding a certain critical value, however, the double peaks
emerge. The higher frequency, the lower is the critical value.
The increasing disorder reduces the band gap effect and smears LE.
We also ploted LE-variance relations. However, with increasing
disorder, we did not observe linear dependence between LE and its
variance, as expected from the single parameter scaling theory.

Now we turn to the discussion of the three chosen models.
When randomness $\Delta=0$, the layered systems become periodic.
Eigenfunctions of wave equation in a periodic environment
are Bloch waves. As is well known, band structure appears
in this situation. The understanding of band structure is very
important and helpful in the following discussions.
The dispersion relation of the Bloch waves in the underlying
periodic system is given by
\beq \cos Kd=\cos k_{a}a\cos k_{b}b-\cosh 2\eta
\sin k_{a}a\sin k_{b}b
\eeq
with $d=a+b$ representing the thickness of the space period
and $K$ the Bloch wave number.
Here for model 1 and 2 function $\cosh 2\eta$ is defined as
\beq
\cosh 2\eta=\frac{1}{2}\left(\sqrt\frac{\epsilon_{b}}{\epsilon_{a}}+
\sqrt\frac{\epsilon_{a}}{\epsilon_{b}}\right)
\eeq
and for model 3 it is given by
\beq
\cosh 2\eta=\frac{1}{2}\left(gh+
\frac{1}{gh}\right), \;\;\;\;g=\rho_a/\rho_b,\;\;\;\;h=c_a/c_b.
\eeq
In the frequency ranges that $Kd$ are real,
the waves are freely propagating in the media
and by definition the frequency ranges correspond to the pass bands.
Beyond the pass bands $Kd$ are not purely real numbers, solutions of wave
equation that satisfy appropriate boundary conditions in both the left and
right infinity do not exist, and thus the frequency ranges are
refer to as the band gaps. If the media is semi-infinite and
the periodicity is ended
by a boundary, say, the left boundary, then the waves are localized
in the vicinity of the boundary. The penetration depth is equal to
$1/|{\rm Im}(K)|$.
Fig.~1(a1)-(c1) plot the band structures of model 1-3.
Solid curves represent ${\rm Re}(Kd)$ and cover the pass
bands. Broken lines cover the band gaps and represent
the inverse penetration depths ${\rm Im}(Kd)$.
Model 1 has very wide pass bands and very narrow
band gaps. Band gaps are wider than pass bands in model 2.
Model 3 has very wide band gaps and very narrow pass bands.
An important feature of model 3 is that in the band gaps
the the penetration depth is very small. Even in the first gap
(which has the longest penetration depth) the penetration depth
is smaller than one period.

To see how randomness influences the transmission properties
we select the frequency range that around the second gap and
plot the transmission curves on (a2)-(c2). In this calculation
1600 layers (800 periods) are used in model 1 and 200 layers
(100 periods) are used in model 2 and 3. In the gaps the transmission
rate is almost zero. When randomness is
increased we observe that the transmission rate in the pass
bands have been reduced much in the model 2 and model 3. On the
other hand, the band gaps in model 2 and 3 seem to be more
robust than in model 1.

To further explore the influence of randomness we study both
LE and its variance.
The results are shown in Fig.~\ref{figure2}.
(A1)-(A5) are results of model 1 and (B1)-(B5) are results of model 2.
LE is denoted as $\gamma$ and its variance
is denoted as $\mbox{var}(\gamma)$. Here $\gamma$ and
$\mbox{var}(\gamma)$ are defined as
\beq
\gamma=\lim_{N\rightarrow \infty}\langle\gamma_N\rangle
\eeq
with
\beq
\gamma_N=\frac{1}{2N}\ln \left(\frac{1}{T_{N}}\right)
\eeq
and
\beq
{\rm var}(\gamma)=\lim_{N\rightarrow \infty}
(\langle\gamma^2_N\rangle-\langle\gamma_N\rangle^2),
\eeq
here $T_N$ is the transmission rate for system with $2N$ layers
($N$ periods in the corresponding periodic system), and
notation $\langle\cdots\rangle$ represents the ensemble average.
The sample size is chosen
in such a way that it is much larger than the localization
length and the ensemble average is carried out over 200 random
configurations.
As expected, when the randomness is small the LE can be approximated
by the inverse penertration depth for wave propagating in the
underlying semi-infinite periodic system. Like in the case
discussed in Ref.\cite{Deych1}, the double peaks of
$\mbox{var}(\gamma)$ first appear near the gap edges.
The peaks of $\mbox{var}(\gamma)$ imply the
fluctuations of transmission. Further increasing $\Delta$,
the $\mbox{var}(\gamma)$ peaks become
fatter and flatter. We observe that if in the vicinity of a peak
there is another peak, then increasing the randomness
will cause them to merge.
For model 1, as $\Delta$ is increased, a pair of
$\mbox{var}(\gamma)$ peaks merge with each other inside the gap,
following the scenario of Ref\cite{Deych1}.
However, for model 2 when
the randomness is increased the pairs of $\mbox{var}(\gamma)$ peaks
tend to merge with each other in the pass bands and finally destroy
the pass bands.
It seems that in model 2 the merging of the double peaks of
$\mbox{var}(\gamma)$ in a pass band will never be
completed. The merging tendency of a pair of double peaks merely
increase the LE and destroy the pass band they belong to.

We also plot the LE versus its variance in
Fig.~\ref{figure3}.
There we observe that although for model 1
(with small dielectric contrast between two layers like
that studied in \cite{Deych1}) SPS seems still a good approximation
in large disorder limit,
deviation from SPS is clearly observed in model 2. Similar quantities
have also been calculated for model 3 and ploted in
Fig.~\ref{figure4} and Fig.~\ref{figure5}. There we also observe
novel behaviors of $\gamma$ and $\mbox{var}(\gamma)$ and
find even larger
deviation from SPS as reported in \cite{LY}.

From these observations, we find:

\begin{itemize}
  \item How does the $\mbox{var}(\gamma)$ vary with the randomness
  depends on many parmeters. For example, the impedance
  constrast, the thickness ratio and which kind of layers (for example,
  high phase velocity, low phase velocity etc.) is randomized.
  \item When impedance constrast is large, it seems that the deviation
  from SPS is a usual feature. The results reported in \cite{Deych1} is
  based more or less on the fact that the impedance contrast between two
  layers is closed to 1.
\end{itemize}

In order to understand why the large deviation from SPS in model 3
is established we study its transmission properties in more detail.
In Fig.~\ref{figure6}(a) we plot the transmission rate of the AC waves
through a pair of air blocks of thickness $a$. The two air blocks
are seperated by a water layer of thickness $b=9999a$. Similarly
in Fig.~\ref{figure6}(b) we plot the transmission for the system
with 100 air blocks. Comparing these two diagrams we find that
the air blocks are very strong scatterers and thus as few as only
two air blocks are enough to determine the ranges
of band gaps and pass bands.
If the localization effect
of AC waves in model 3 is mainly determined by the multiple sacttering
of AC waves between pairs of air blocks, then one would expect
that for a system with 100 air blocks
the transmission $T_{N=100}$ in the band gaps can be approximated
by $T^{50}_{N=2}$, here $T_{N=2}$ refers to the
transmission rate for two air blocks illustrated in \ref{figure6}(a).
We indeed observed this result in
Fig.~\ref{figure6}(c). When randomness is not very large, this local
effect explains why the phase averaging process in model 3 is
so inefficient that the randomness cannot modify the LE
much in the band gaps.
\section{Concluding Remarks}
In this paper we studied the statistics of
localization properties in one-dimensional
layered systems.
Luapunov exponent and its variance are
compared for three chosen models. We find that
the band structures of the corresponding periodic
systems influence the localization properties much
if the impedance contrast between neighboring layers
is not close to 1. In general the single parameter scaling is not very
accurate and more model-dependent parametrs should be included
in the detailed discriptions of localization behaviors.

\section*{ACKNOWLEDGEMENT}
The work received support from National Science Council (No.
NSC89-2611-M008-002 and NSC89-2112-M008-008).


\input{epsf}
\begin{figure}[hbt]
\begin{center}
\epsfxsize=3in \epsffile{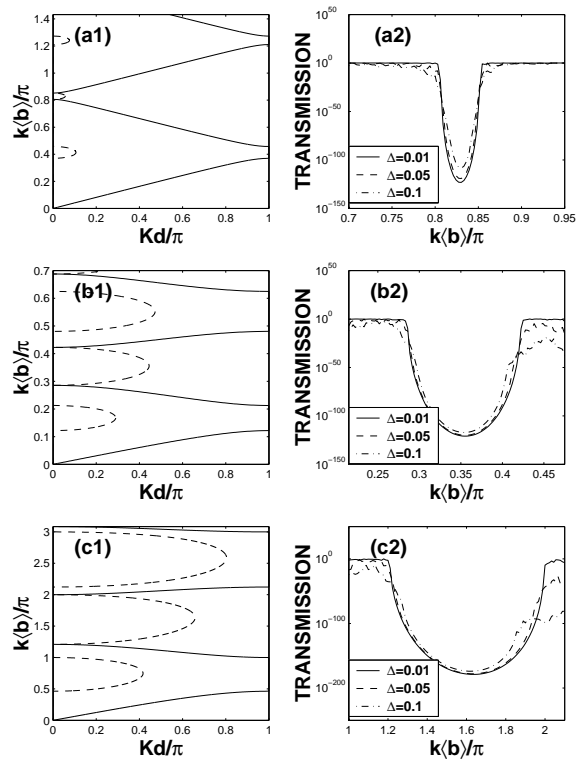}\vspace{5mm}
\caption{\label{figure1}\small Band structures (solid lines) and
transmission curves for the three models disscussed in this paper.
(a1) and (a2) are for the first model. (b1), (b2) are for the
second model. (c1) and (c2) are for the third model.
The broken lines in (a1)-(c1) represent the inverse penetration
depth $\mbox{Im}(K)$.}
\end{center}
\end{figure}
\newpage
\input{epsf}
\begin{figure}[hbt]
\begin{center}
\epsfxsize=3in \epsffile{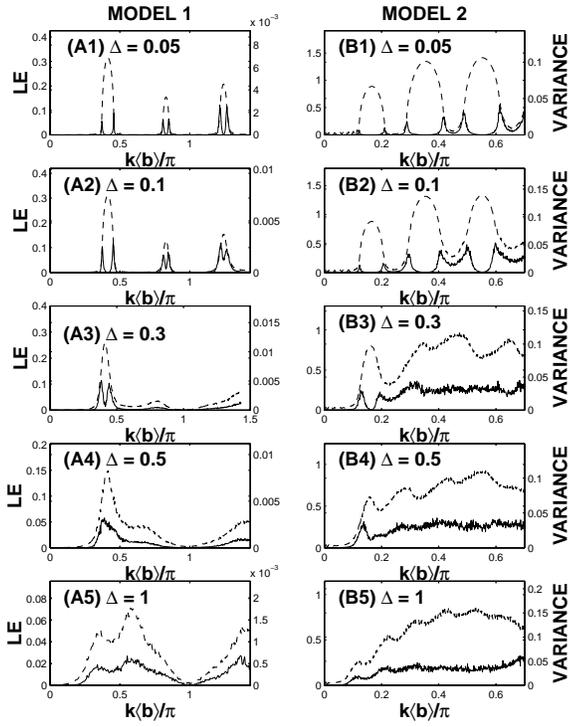}\vspace{5mm}
\caption{\label{figure2}\small LE (broken lines) and its variance
(solid lines) for the first ((A1)-(A5)) and the second
((B1)-(B2)) models.}
\end{center}
\end{figure}
\input{epsf}
\begin{figure}[hbt]
\begin{center}
\epsfxsize=3in \epsffile{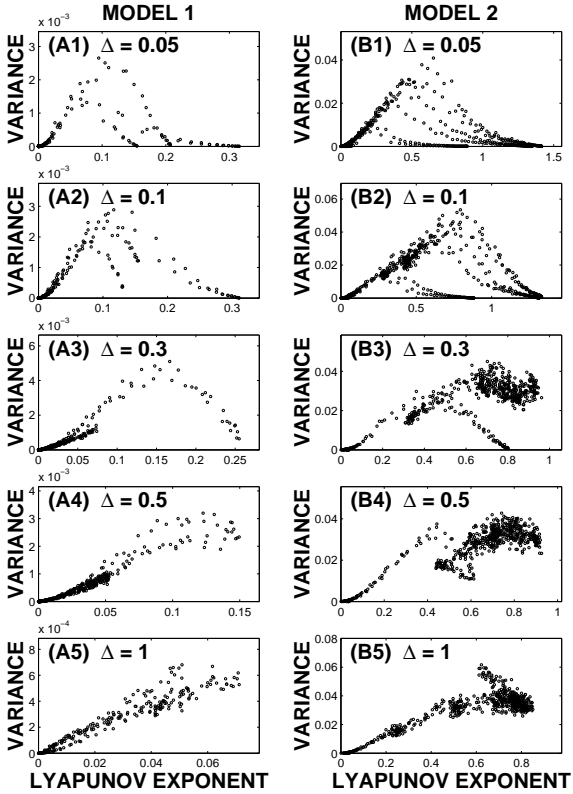}\vspace{5mm}
\caption{\label{figure3}\small var($\gamma$) versus $\gamma$
for the first and the second models.}
\end{center}
\end{figure}

\newpage
\input{epsf}
\begin{figure}[hbt]
\begin{center}
\epsfxsize=3in \epsffile{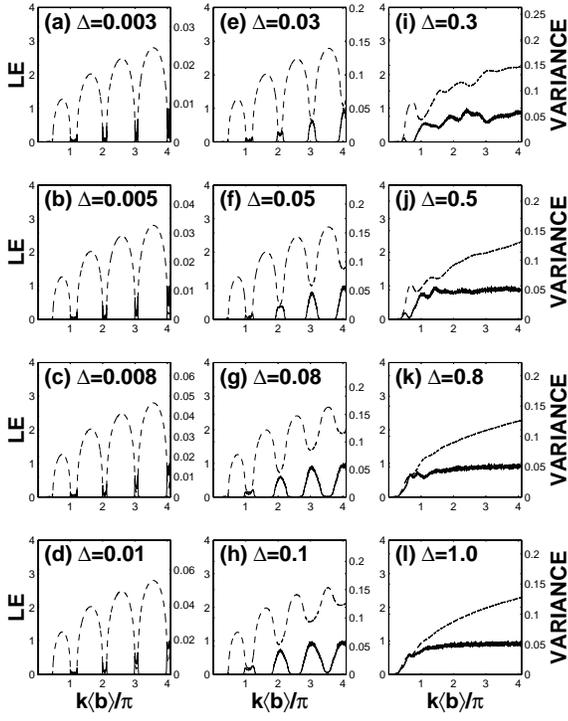}\vspace{5mm}
\caption{\label{figure4}\small Lyapunov exponent (broken lines)
and its variance (solid lines) for the third model.}
\end{center}
\end{figure}

\input{epsf}
\begin{figure}[hbt]
\begin{center}
\epsfxsize=3in \epsffile{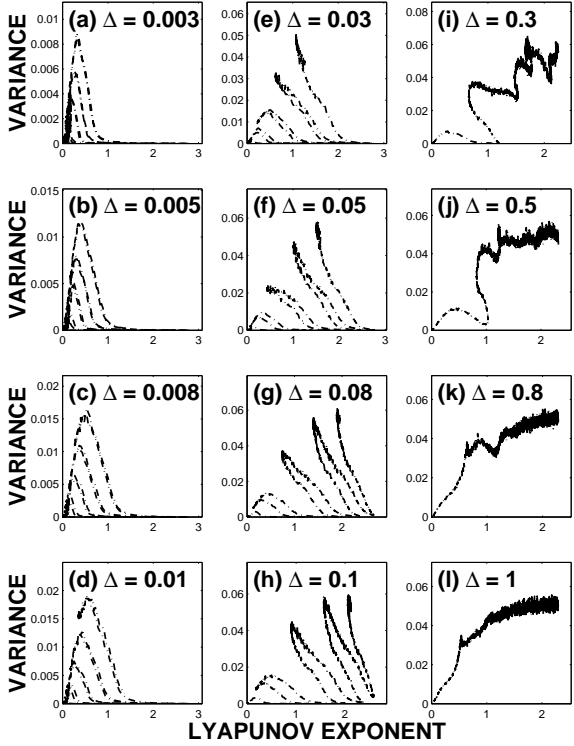}\vspace{5mm}
\caption{\label{figure5}\small var($\gamma$) versus $\gamma$
for the third model.}
\end{center}
\end{figure}

\newpage
\input{epsf}
\begin{figure}[hbt]
\begin{center}
\epsfxsize=3in \epsffile{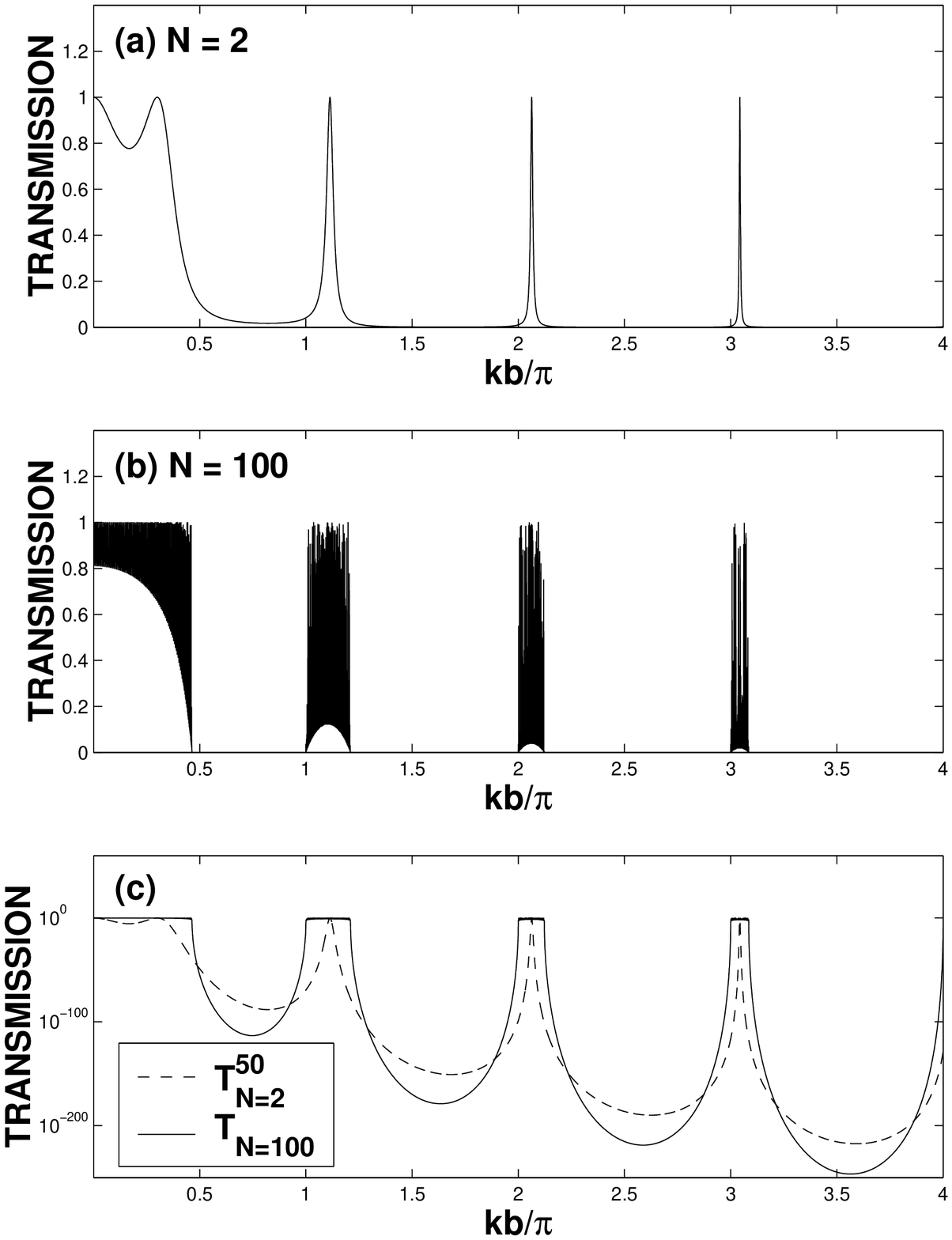}\vspace{5mm}
\caption{\label{figure6}\small Transmission vs dimensionless frequency
for the third model. (a) Transmission for the N=2 system
(two air blocks). (b) Transmission for the N=100 system (100
air blocks). (c) Comparison
between $T_{N=2}^{50}$ (broken line) and $T_{N=100}$ (solid line).}
\end{center}
\end{figure}

\begin{references}
\bibitem{Anderson} P. W. Anderson, Phys. Rev. {\bf 109}, 1492
(1958).

\bibitem{Baluni} V. Baluni and Willemsen, Phys. Rev. {\bf A} 31,
3358 (1985).

\bibitem{SEGC} C. M. Soukoulis, E. N. Economou, G. S. Grest and M. H. Cohen, Phys. Rev. Lett.
 {\bf 62}, 575 (1989).

\bibitem{Sor} D. Sornette and O. Legrand, J. Acoust. Soc. Am. {\bf
92} (1992).

\bibitem{Maradudin93} A. R. McGurn, K. T. Christensen, F. M. Mueller,
and A. A. Maradudin, Phys. Rev. B {\bf 47}, 13120 (1993).

\bibitem{Ye1} Z. Ye and A. Alvarez, Phys. Rev. Lett. {\bf 80},
3503 (1998).

\bibitem{Ye2} Z. Ye, H. Hsu, E. Hoskinson, and A. Alvarez, Chin. J.
Phys. {\bf 37}, 343 (1999).

\bibitem{SIAM} M. Asch, W. Kohler, G. Papanicolaou, M. Postel, and
B. White, SIAM Review {\bf 33}, 519-625 (1991).

\bibitem{Hodges83} C. H. Hodges and J. Woodhouse, J. Acoust. Soc.
Am. {\bf 74}, 894 (1983).

\bibitem{Sch} R. Dalichaouch, J. P. Armstrong, S. Schultz, P. M. Platzman and
 S. L. McCall, Nature, {\bf 354}, 53 (1991).

\bibitem{TAA} P. W. Anderson, D. J. Thouless, E. Abrahams, and
D. S. Fisher, Phys. Rev. B {\bf 22}, 3519 (1980).

\bibitem{Sak} J. Sak and B. Kramer,
Phys. Rev. B {\bf 24}, 1761 (1981).

\bibitem{Mello} J. Flores, P. A. Mello, and G. Monsiv$\acute{a}$is,
Phys. Rev. B {\bf 35}, 2144 (1987).

\bibitem{Shapiro} Avraham Cohen, Yehuda Roth, and Boris Shapiro,
Phys. Rev. B {\bf 38}, 12125 (1988).

\bibitem{Deych1} L. I. Deych, D. Zaslavsky and A. A. Lisyansky,
Phys. Rev. Lett. {\bf 81}, 5390 (1998)
\bibitem{Deych2} L. I. Deych, A. A. Lisyansky,
and B. L. Altshuler, Phys. Rev. Lett. {\bf 84}, 2678 (2000).
\bibitem{LY} Pi-Gang Luan and Zhen Ye, Phys. Rev. E {\bf 63}, 066611
(2001).

\end{references}
\end{document}